\documentclass[fleqn,11pt]{article}
\usepackage{amsmath,amssymb}
\usepackage{xcolor}
\usepackage{verbatim}
\usepackage{amsfonts,amssymb,cite}
\usepackage{graphicx}
\def\noi{\noindent}

\newcommand{\Title}[1]{\noi {{\Large\bf #1}}\\[1ex]}

\newcommand{\Author}[2]{\noi{\bf #1}\\[2ex]\noi{\normalsize\it #2}\\}

\newcommand{\Abstract}[1]{\vskip 2mm \begin{center}
        \parbox{16.4cm}{\small\noi #1} \end{center}\medskip}
\newcommand{\foom}[1]{\protect\footnotemark[#1]}

\def\nqq{\hspace*{-2em}}

\def\Jl#1#2{#1 {\bf #2},\ }

\def\ApJ#1 {\Jl{Astroph. J.}{#1}}
\def\CQG#1 {\Jl{Class. Quantum Grav.}{#1}}
\def\DAN#1 {\Jl{Dokl. AN SSSR}{#1}}
\def\GC#1 {\Jl{Grav. Cosmol.}{#1}}
\def\GRG#1 {\Jl{Gen. Rel. Grav.}{#1}}
\def\JETF#1 {\Jl{Zh. Eksp. Teor. Fiz.}{#1}}
\def\JETP#1 {\Jl{Sov. Phys. JETP}{#1}}
\def\JHEP#1 {\Jl{JHEP}{#1}}
\def\JMP#1 {\Jl{J. Math. Phys.}{#1}}
\def\NPB#1 {\Jl{Nucl. Phys. B}{#1}}
\def\NP#1 {\Jl{Nucl. Phys.}{#1}}
\def\PLA#1 {\Jl{Phys. Lett. A}{#1}}
\def\PLB#1 {\Jl{Phys. Lett. B}{#1}}
\def\PRD#1 {\Jl{Phys. Rev. D}{#1}}
\def\PRL#1 {\Jl{Phys. Rev. Lett.}{#1}}

\def\lal{&&\nqq {}}

\def\beq{\begin{equation}}
\def\eeq{\end{equation}}
\def\bear{\begin{eqnarray}}
\def\bearr{\begin{eqnarray} \lal}
\def\ear{\end{eqnarray}}
\def\earn{\nonumber \end{eqnarray}}

\newcommand{\MI}{$\mathfrak{M}_1$}

\newcommand{\Fig}[3]{%
\begin{center}
\parbox{8cm}{%
\refstepcounter{figure}\includegraphics[width=8cm,height=#2cm]{#1} \noindent Fig. \thefigure:\quad
#3}\end{center}}
\newcounter{strochka}
\newcommand{\stroka}[1]{\refstepcounter{strochka}\par\noindent\textsl{\Roman{spisok}.\arabic{strochka}}. \; \textsl{#1}}
\newcounter{spisok}
\begin{document}
\thispagestyle{empty}
\twocolumn[

\vspace{1cm}

\Title{Two-field model of gravitational-scalar instability and the formation of supermassive black holes in the early Universe\foom 1}

\Author{Yu. G. Ignat'ev}
    {Institute of Physics, Kazan Federal University, Kremlyovskaya str., 16A, Kazan, 420008, Russia}


\Abstract
 {Based on the previously formulated mathematical model of a statistical system with scalar interaction of fermions and the theory of gravitational-scalar instability of a cosmological model based on a two-component statistical system of scalarly charged degenerate fermions, a numerical model of the cosmological evolution of gravitational-scalar perturbations in the presence of classical and phantom scalar fields is constructed and studied. The gravitational-scalar instability in the early stages of expansion in the model under study arises at sufficiently large scalar charges, and the instability develops near the unstable points of the vacuum doublet.
 Shortwave perturbations of the free phantom field turn out to be stable at stable singular points of the vacuum doublet. It is shown that for sufficiently large scalar charges, mass perturbations can grow to the values of masses Black Hole Seeds (BHS).
}
\bigskip

] 
\section{Introduction}
%
%
Simulation of the evolution of supermassive black holes (SMBHs) with mass $\sim10^9\div10^{10}M_\odot$ observed in quasars at redshifts $z\gtrsim 6$ \cite{Fan} showed that for their formation by the corresponding moment of cosmological time (t=0.6 Myr. years) already at $z\sim 7$ supermassive black hole nuclei (BHS) with mass \cite{Zhu}
\begin{equation}\label{M_nc}
M_{bhs}\sim 10^4\div 10^6 M_\odot\approx 10^{42}\div10^{44}m_{\mathrm{pl}}.
\end{equation}
Due to the fact that the standard mechanisms of gaseous accretion are not able to provide such a rapid formation of objects with a mass \eqref{M_nc}, \cite{GC_21_1} -- \cite{GC_21_2} studied the short-wavelength instability of a cosmological model based on a one-component a system of degenerate fermions with a scalar Higgs interaction, either classical or phantom.
It was also shown that in the \emph{hard WKB approximation}
\begin{eqnarray}\label{WKB0}
n\eta\gg 1,\\
\label{stiff-WKB}
n^2\gg a^2 m^2,
\end{eqnarray}
where $n$ is the wave number of perturbations in the spatially flat Friedmann metric
\begin{eqnarray}\label{ds0}
ds_0^2=dt^2-a^2(t)(dx^2+dy^2+dz^2)\equiv\nonumber\\
a^2(\eta)(d\eta^2-dx^2-dy^2-dz^2),\;(t=\int ad\eta)
\end{eqnarray}
and $m$ is the mass of quanta of the scalar field, a system with a classical scalar interaction may turn out to be unstable, while a system with a phantom scalar interaction is stable.  Further, in \cite{Ign22_I}, \cite{Ign22_II} a model of BHS formation was proposed based on the \emph{gravi\-ta\-ti\-onal - scalar instability} mechanism of a one-component system of degenerate fermions with the classical Higgs interaction. In this case, the rigidity condition for the WKB approximation was removed. In contrast to the standard hydrodynamic instability, the gravi\-ta\-tional-scalar instability deve\-lops exponentially fast. The investigations carried out in \cite{Ign22_II} showed the fundamental possibility of early formation of BHS based on the me\-cha\-nism of gravitational-scalar instability. At the same time, however, a number of issues related to this me\-cha\-nism remained unexplored. Such questions, in particular, include the question of the influence of the phantom scalar interaction on the development of instability.

Earlier, in the paper \cite{Ign_GC21_Un} the stability of a two-component system of degenerate fermions with scalar classical and phantom charges was partially investigated. However, this study was not complete, since it contained some numerical examples, but a sufficiently serious analysis of the influence of the phantom component on both the instability parameters and the parameters of the BHS formation mechanism was not carried out. In addition, in this work, the approximation of small scalar charges, which can be violated at sufficiently long evolution times, was considered. This approximation made it possible to analytically calculate the perturbation increments. The present study makes up for these shortcomings. In this case, removing the conditions for the smallness of scalar charges, we will investigate the growth rates of perturbations by numerical methods. In addition, we will pay attention to the question of the stability of Higgs classical and phantom scalar fields.

\section{Mathematical model of the cos\-mological system of degenerate scalarly charged fermions}
Bearing in mind the application of a number of results of this article in its second part, we first consider the general mathematical model \MI\ for the case of an asymmetric scalar doublet represented by the canonical scalar field $\Phi$
and a phantom scalar field $\varphi$. The cosmological model for the canonical scalar singlet $\Phi$ will be considered as a special case of the general \MI\ model under the passage of a number of its parameters to the limit.
\subsection{Self-consistent system of equations for degenerate scalarly charged\newline fermions}
Let us consider a cosmological model based on a self-gravitating two-component system of singly scalarly charged degenerate fermions interacting through a pair of scalar Higgs fields, canonical, $\Phi$, and phantom, $\varphi$. This model is described, firstly, by the system of Einstein equations
\begin{equation}\label{Eq_Einst_G}
R^i_k-\frac{1}{2}\delta^i_k R=8\pi T^i_k+ \delta^i_k \Lambda_0,
\end{equation}
where
\[T^i_k=T^i_{(s)k}+T^i_{(p)k},\]
$T^i_{(s)k}$ is the energy-momentum tensor of scalar fields
\begin{eqnarray}\label{T_s}
T^i_{(s)k}=\frac{1}{16\pi }\bigl(2\Phi^{,i}\Phi_{,k} -\delta^i_k\Phi_{,j} \Phi{,j}
 +2V_z(\Phi)\delta^i_k \bigr)\nonumber\\
 -\frac{1}{16\pi }\bigl(2\varphi^{,i}\varphi_{,k} -\delta^i_k\varphi_{,j} \varphi{,j}
 -2V_\zeta(\varphi)\delta^i_k \bigr),
\end{eqnarray}
and
\begin{eqnarray}
\label{Higgs}
V_z(\Phi)=-\frac{\alpha}{4} \left(\Phi^{2} -\frac{m^{2} }{\alpha}\right)^{2};\nonumber\\
V_\zeta(\varphi)=-\frac{\beta}{4} \left(\varphi^{2} -\frac{\mathfrak{m}^{2}}{\beta}\right)^{2}
\end{eqnarray}
are the potential energies of the corresponding scalar fields, $\alpha $ and $\beta$ are their self-action constants, $m$ and $\mathfrak{m}$ are their quantum masses. As a carrier of scalar charges, we consider a two-component degenerate system of fermions, in which the cano\-nical charge carriers $z$-fermions have the canonical charge $e_z$ and the Fermi momentum $\pi_{(z)}$, and the carriers of the phantom charge are $\zeta$-fermions have phantom charge $e_\zeta$ and Fermi momentum $\pi_{(\zeta)}$. The dynamic masses of these fermions in the case of zero bare masses are \cite{TMF_Ign_Ign21}
\begin{equation}\label{m_*(pm)}
m_{z}=e_z\Phi;\qquad m_{\zeta}=e_\zeta\varphi.
\end{equation}

The seed cosmological constant $\Lambda_0$, which ap\-pears in the right-hand side of the Einstein equations \eqref{Eq_Einst_G}, is related to its observed value $\Lambda$ by the rela\-ti\-onship
\[
\Lambda=\Lambda_0-\frac{1}{4}\sum\limits_r \frac{m^4_r}{\alpha_r}.\]

Further, the energy-momentum tensor of an equilibrium statistical system is equal to:
\begin{equation}\label{T_p}
T^i_{(p)k}=(\varepsilon_p+p_p)u^i u_k-\delta^i_k p_p,
\end{equation}
where $u^i$ is the macroscopic velocity vector of the statistical system, $\varepsilon_p$ and $p_p$ are its energy density and pressure. These macroscopic scalars, as well as other scalar functions that determine the macroscopic characteristics of the statistical system, are equal for a two-component statistical system of degenerate fermions (see, for example, \cite{TMF_Ign_Ign21}):
\begin{equation}\label{2_3}
n^{(a)}=\frac{1}{\pi^2}\pi_{(a)}^3;
\end{equation}
\begin{eqnarray}
\label{2_3a_2}
\varepsilon_p=\frac{e^4_z \Phi^4}{8\pi^2}F_2(\psi_z)+\frac{e^4_\zeta \varphi^4}{8\pi^2}F_2(\psi_\zeta);\\
\label{2_3b_2}
p_p  =\displaystyle \frac{e^4_z \Phi^4}{24\pi^2}(F_2(\psi_z)-4F_1(\psi_z))+\nonumber\\
\frac{e^4_\zeta \varphi^4}{24\pi^2}(F_2(\psi_\zeta)-4F_1(\psi_\zeta));\\
\label{2_3c}
\displaystyle
\sigma^z=\frac{e_z^4 \Phi^3}{2\pi^2}F_1(\psi_z);\;\sigma^\zeta=\frac{e_\zeta^4 \varphi^3}{2\pi^2}F_1(\psi_\zeta),
\end{eqnarray}
where the macroscopic scalars $n_{(a)}$ -- \emph{scalar particle number density} and $\sigma^z$ and $\sigma^\zeta$ -- \emph{scalar charge densities} $e_z$ and $e_\zeta$,
\begin{equation}\label{psi_zzeta}
\psi_z=\frac{\pi_{(z)}}{|e_z\Phi|}; \qquad \psi_\zeta=\frac{\pi_{(\zeta)}}{|e_\zeta\varphi|},
\end{equation}
and also to shorten the letter, the functions $F_1(\psi)$ and $F_2(\psi)$ are introduced:
\begin{eqnarray}\label{F_1}
F_1(\psi)=\psi\sqrt{1+\psi^2}-\ln(\psi+\sqrt{1+\psi^2});\\
\label{F_2}
F_2(\psi)=\psi\sqrt{1+\psi^2}(1+2\psi^2)\nonumber\\
-\ln(\psi+\sqrt{1+\psi^2}).
\end{eqnarray}
In addition, we write down the expression we need below for \emph{densities of scalar charges}, $\rho_{(a)}$, which are determined using the charge number density $n_{(a)}$ \cite{Ignat21_TMP} and not coincide in general
with the scalar charge densities $\sigma^z$ and $\sigma^\zeta$ introduced above:
\begin{equation}\label{rho}
\rho_{(a)}=e_{(a)}n_{(a)}=\frac{e_{(a)}\pi^3_{(a)}}{\pi^2}.
\end{equation}
Finally, the equations of scalar fields for the system under study take the form:
\begin{eqnarray}\label{Box(Phi)=sigma_z}
\Box \Phi + m^2\Phi-\alpha\Phi^3 = -\frac{4}{\pi^2}e^4_z\Phi^4 F_1(\psi_z),\\
\label{Box(varphi)=sigma_zeta}
-\Box \varphi + \mathfrak{m}^2\varphi-\beta\varphi^3 = -\frac{4}{\pi^2}e^4_\zeta\varphi^4 F_1(\psi_\zeta).
\end{eqnarray}
Note that the scalar charge densities $\sigma^z$ and $\sigma^\zeta$ \eqref{2_3c} are the sources of the corresponding scalar fields $\Phi$ and $\varphi$, while the scalar charge densities $\rho_z $ and $\rho_\zeta$
are defined by a prime number of charges and, unlike \eqref{2_3c}, are related to the corresponding \emph{conserved scalar charges}
\begin{equation}\label{Q}
Q_{(a)}=\int \rho_{(a)}dV.
\end{equation}
\subsection{Background state for the cosmolo\-gical model \MI}
Let us further consider the space-flat model of the Friedman universe \eqref{ds0}. A strict consequence of the general relativistic kinetic theory for statistical systems of completely degenerate fermions is the Fermi momentum conservation law $\pi_{(a)}$ for each component
 \begin{equation}\label{ap}
 a(t)\pi_{(a)}(t)=\mathrm{Const}.
 \end{equation}
Assuming in what follows, for definiteness, $a(0)=1$ and
\begin{eqnarray}\label{a-xi}
\xi=\ln a;\quad \xi\in(-\infty,+\infty); \quad \xi(0)=0,\\
\label{psi0}
\pi_{(z)}=\pi_c \mathrm{e}^{-\xi},\; \pi_{(\zeta)}=\pi_f \mathrm{e}^{-\xi},\nonumber\\
 (\pi_c=\pi_{(z)}(0),\pi_f=\pi_{(\zeta)}(0)),
\end{eqnarray}
write out the complete normal system of Einstein equations and scalar fields $\Phi(t)$ and $\varphi(t)$ for this two-component system of scalarly charged degenerate fermions\cite{TMF_Ign_Ign21 } in an apparently non-singular form:
\begin{eqnarray}
\label{dxi/dt-dPhi_Phi}
\dot{\xi}=H;\qquad \dot{\Phi}=Z;\qquad \dot{\varphi}=z;\\
\label{dH/dt_M1}
\dot{H}=-\frac{Z^2}{2}+\frac{z^2}{2}-\frac{4\mathrm{e}^{-3\xi}}{3\pi}\times\nonumber\\
\biggl(\pi_c^3\sqrt{\pi_c^2\mathrm{e}^{-2\xi}+e^2\Phi^2}+\pi_f^3\sqrt{\pi_f^2\mathrm{e}^{-2\xi}+\epsilon^2\varphi^2}\biggr);
\end{eqnarray}
\begin{eqnarray}
\label{dZ/dt_M1}
\dot{Z}=-3HZ-m^2\Phi+\alpha\Phi^3-\nonumber\\[6pt]
\frac{4e^2\pi_c\mathrm{e}^{-\xi}}{\pi}\Phi\sqrt{\pi^2_c \mathrm{e}^{-2\xi}+e^2\Phi^2}+\nonumber\\
\frac{4e^4}{\pi}\Phi^3\ln\biggl(\frac{\pi_c\mathrm{e}^{-\xi}+\sqrt{\pi^2_c \mathrm{e}^{-2\xi}+e^2\Phi^2}}{|e\Phi|} \biggr);
\end{eqnarray}
\begin{eqnarray}
\label{dzZ/dt_M1}
\dot{z}=-3Hz+\mathfrak{m}^2\varphi-\beta\varphi^3+\nonumber\\[6pt]
\frac{4\epsilon^2\pi_f\mathrm{e}^{-\xi}}{\pi}\varphi\sqrt{\pi^2_f \mathrm{e}^{-2\xi}+\epsilon^2\varphi^2}-\nonumber\\
\frac{4\epsilon^4}{\pi}\varphi^3\ln\biggl(\frac{\pi_f\mathrm{e}^{-\xi}+\sqrt{\pi^2_f \mathrm{e}^{-2\xi}+\epsilon^2\varphi^2}}{|\epsilon\varphi|} \biggr).
\end{eqnarray}
The system of equations \eqref{dxi/dt-dPhi_Phi} -- \eqref{dzZ/dt_M1} has as its first integral the total energy integral \cite{TMF_Ign_Ign21}, which can be used to determine the initial value of the function $H(t)$
\begin{eqnarray}
\frac{Z^2}{2}+\frac{z^2}{2}-\frac{m^2\Phi^2}{2}+\frac{\alpha\Phi^4}{4}-\frac{\mathfrak{m}^2\varphi^2}{2}+\frac{\beta\varphi^4}{4} \nonumber\\
-\frac{e^{-\xi}}{\pi}\biggl(\pi_c\sqrt{\pi_c^2\mathrm{e}^{-2\xi}+e^2\Phi^2}\bigl(2\pi^2_c\mathrm{e}^{-2\xi}+e^2\Phi^2\bigr)\nonumber\\
+\pi_f\sqrt{\pi_f^2\mathrm{e}^{-2\xi}+\epsilon^2\varphi^2}\bigl(2\pi^2_f \mathrm{e}^{-2\xi}+\epsilon^2\varphi^2\bigr)\biggr)\nonumber\\
%
\label{SurfEinst_M1}
+\frac{e^4\Phi^4}{\pi}\ln\biggl(\frac{\pi_c\mathrm{e}^{-\xi}+\sqrt{\pi^2_c \mathrm{e}^{-2\xi}+e^2\Phi^2}}{|e\Phi|} \biggr)\nonumber\\
+\frac{\epsilon^4\varphi^4}{\pi}\ln\biggl(\frac{\pi_f\mathrm{e}^{-\xi}+\sqrt{\pi^2_f \mathrm{e}^{-2\xi}+\epsilon^2\varphi^2}}{|\epsilon\varphi|} \biggr)\nonumber\\
+3H^2-\Lambda=0.
\end{eqnarray}

The \MI\ models were studied in \cite{GC_21_3}, where it is shown that such cosmological models, firstly, have an initial singularity with a short-term ultrarelativistic expansion phase passing to the inflation mode and, secondly, depending on the value cosmological constant can also have a finite singularity.

We introduce the invariant characteristics of the unperturbed cosmological model, which are necessary in what follows, \emph{invariant cosmological acceleration}
\begin{equation}\label{Omega}
\Omega=1+\frac{\dot{H}}{H^2}
\end{equation}
and \emph{invariant curvature of four-dimensional space} $K$
\begin{eqnarray}\label{K}
K\equiv\sqrt{R_{ijkl}R^{ijkl}}=H^2\sqrt{6(1+\Omega^2)}\nonumber\\
\equiv \sqrt{6}\sqrt{H^4+\bigl(H^2+\dot{H}\bigr)^2}\geq0.
\end{eqnarray}

\section{Gravitational-scalar instability of the \MI\ model in the short-wavelength limit for a two - com\-ponent system}
\subsection{WKB approximation of instability theory}
For the case of an asymmetric scalar Higgs doublet and a two-component system of singly scalarly charged degenerate fermions, we formulate the main results of \cite{STFI_20} (see also \cite{Ign_GC21_Un}), in which the evolution of gravitational-scalar perturbations of the \eqref{ds0} metric and scalar fields in the \MI model, for the case of purely longitudinal perturbations of the metric \eqref{ds0} in the form \cite{Lifshitz}
\begin{eqnarray}
\label{metric_pert}
ds^2=ds^2_0-a^2(\eta)h_{\alpha\beta}dx^\alpha dx^\beta,
\end{eqnarray}
where $ds_0$ is the unperturbed spatially flat Friedmann metric \eqref{ds0} in conformally flat form
and, for definiteness, the wave vector is directed along the $Oz$ axis:
\begin{eqnarray}\label{nz1}
 h_{11}=h_{22} =\frac{1}{3}[\lambda(t)+\frac{1}{3}\mu(t)]\mathrm{e}^{inz};\nonumber\\
\label{nz13}
h=\mu(t)\mathrm{e}^{inz};\; h_{12}=h_{13}= h_{23}=0;\nonumber\\
\label{nz2}
h_{33}=\frac{1}{3}[-2\lambda(t)+\mu(t)]\mathrm{e}^{inz}.
\end{eqnarray}
At the same time, the matter in the \MI\ model in the case of an asymmetric scalar Higgs doublet and a two - component system of degenerate scalarly charged fermions is completely determined by four scalar functions -- $\Phi(z,\eta)$, $\varphi(z,\eta)$, $ \pi_{(z)}(z,\eta)$ and $\pi_{(\zeta)}(z,\eta)$, as well as the velocity vector $u^i(z,\eta)$. Let us expand these functions into a series in terms of the smallness of perturbations with respect to the corresponding functions against the background of the Friedmann metric \eqref{ds0}\footnote{For scalar singlets, see \cite{GC_21_1}. To avoid cumbersome notation, we have retained the notation for the perturbed values of the functions, distinguishing them only by arguments.}
\begin{eqnarray}\label{dF-drho-du}
\Phi(z,\eta)=\Phi(\eta)+\delta\Phi(\eta)\mathrm{e}^{inz};\nonumber\\
\varphi(z,\eta)=\varphi(\eta)+\delta\varphi(\eta)\mathrm{e}^{inz};\nonumber\\
\pi_{(z)}(z,t)=\pi_{(z)}(\eta)(1+\delta(\eta)\mathrm{e}^{inz});\nonumber\\
\pi_{(\zeta)}(z,t)=\pi_{(\zeta)}(\eta)(1+\delta(\eta)\mathrm{e}^{inz});\\
\sigma^z(z,\eta)= \sigma^z(\eta)+\delta\sigma^z(\eta)\mathrm{e}^{inz};\nonumber\\
\sigma^\zeta(z,\eta)= \sigma^\zeta(\eta)+\delta\sigma^\zeta(\eta)\mathrm{e}^{inz};\nonumber\\
u^i=\frac{1}{a}\delta^i_4+\delta^i_3 v(\eta)\mathrm{e}^{inz},\nonumber
\end{eqnarray}
where $\delta\Phi(\eta)$, $\delta\varphi(\eta)$, $\delta(\eta)$, $s_z(\eta),s_\zeta(\eta)$ and $v(\eta)$ are functions of the first order of smallness compared to their unperturbed values, and the relative per\-tur\-ba\-tions of the Fermi momentum $\delta(\eta)$ do not depend on the kind of fermions \cite{STFI_20}.

In \cite{STFI_20} (see also \cite{Ign_GC21_Un}), the evolution of longitudinal gravitational scalar perturbations of the \MI\ model is studied in the short-wavelength and low-charge approximations.
At the same time, in contrast to the works \cite{GC_21_1} -- \cite{GC_21_2}, the condition of the rigid WKB approximation was not imposed in this work, which makes it possible to consider also sufficiently large wavelengths:
\begin{equation}\label{n<=am}
n^2\gtrsim a^2\{ m^2\Phi,\alpha\Phi^3,\mathfrak{m}^2\varphi,\beta\varphi^3\}.
\end{equation}
In accordance with the WKB method, we represent the perturbation functions $f(\eta)$ in the form
\begin{equation}\label{Eiconal}
f=\tilde{f}(\eta) \cdot \mathrm{e}^{i\int u(\eta)d\eta}; \quad (|u\eta|\sim |n\eta| \gg 1),
\end{equation}
where $\tilde{f}(\eta)$ and $u(\eta)$ are functions of the perturba\-tion amplitude and eikonal that vary slightly along with the scale factor.
In this paper, in contrast to \cite{STFI_20}, we will not impose an additional condition for the smallness of the scalar charge used to simplify the dispersion equation
\[n^2\gg  e^4_{(a)},\; a^2\{m^2,\mathfrak{m}^2\}\gg e^4_{(a)},\]
and we will search for its solutions by numerical methods.

The equations for the amplitudes of perturbations in the zero WKB approximation \eqref{WKB0} take the form of a linear homogeneous system of algebraic equations ($\nu=\lambda+\mu$)
\begin{equation}\label{AX=0}
\!\!\!\mathbb{A}\mathbb{X}\equiv \left(
\begin{array}{ccc|c}
 & & & 0 \\
 & \tilde{\mathbf{A}}_3 & & 0 \\
 & & & 0 \\
 \hline
  & & & \\[-6pt]
 0 & 0 & -\frac{n^2}{3} & -u^2 \\
 \end{array}\right)
 \cdot
\left(\begin{array}{c}
\Phi\\ \varphi\\ \nu\\ \lambda\\
\end{array}
\right)=0,
\end{equation}
where
\small{
\begin{eqnarray}\label{A_gamma}
\!\!\!\!\!\tilde{\mathbf{A}}_3=\left[\begin{array}{lll}
 n^2-u^2+\gamma_{11} & \gamma_{12} & \gamma_{13}n^2\\[12pt]
\gamma_{21} & n^2-u^2+\gamma_{22} & \gamma_{23}n^2\\[12pt]
\gamma_{31} & \gamma_{32} & \gamma_{33}n^2-u^2\\
\end{array}\right]\nonumber
\end{eqnarray}}
and introduced the notation:
\begin{eqnarray}
\label{gamma_ik}
\gamma_{11}=+a^2(m^2-3\alpha\Phi^2+8\pi S^z_\Phi) \nonumber\\
\gamma_{12}= -8\pi a^2S^z_\varphi;\; \gamma_{13}= \frac{e^4_z\Phi^3\psi^2_z}{6\pi^2\varepsilon^\delta_p\sqrt{1+\psi^2_z}};\nonumber\\
\gamma_{22}= -a^2(\mathfrak{m}^2-3\beta\varphi^2+8\pi S^\zeta_\varphi);\nonumber\\
\gamma_{21}= 8\pi a^2S^\zeta_\Phi;\;\gamma_{23}=-\frac{e^4_\zeta\varphi^3\psi^2_\zeta}{6\pi^2\varepsilon^\delta_p\sqrt{1+\psi^2_\zeta}};\\
\!\!\!\!\!\!\gamma_{31}= -3a^2[\Phi(m^2-\alpha\Phi^2)-8\pi P^\Phi];\ \gamma_{33}= \frac{1}{3}+\frac{p^\delta_p}{\varepsilon^\delta_p}; \nonumber\\
\gamma_{32}=  -3a^2[\varphi(\mathfrak{m}^2-\beta\varphi^2)-8\pi P^\varphi].   \nonumber
\end{eqnarray}

In turn, the coefficients of the theory of gravitational-scalar instability \cite{STFI_20} included in the formulas \eqref{gamma_ik} and expressed in terms of the basic functions of the unperturbed model \MI\ $a(t)$ and $\Phi(t) $, as well as through the kinetic coefficients $\psi_a(t)$ \eqref{psi_zzeta}, are equal to:
\begin{eqnarray}\label{de_delta}
 \varepsilon_p^\delta=\frac{e_z^4\Phi^4\psi_z^3\sqrt{1+\psi_z^2}
+e_\zeta^4\varphi^4\psi_\zeta^3\sqrt{1+\psi_\zeta^2}}{\pi^2} >0;\nonumber\\
\varepsilon_p^\Phi=\frac{e_z^4\Phi^3}{2\pi^2}F_1(\psi_z);
\varepsilon_p^\varphi=\frac{e_\zeta^4\varphi^3}{2\pi^2}F_1(\psi_\zeta);\nonumber\\
\Delta_\Phi=\frac{\varepsilon_p^\Phi}{ 8\pi \varepsilon_p^\delta};\;
\Delta_\varphi=\frac{\varepsilon_p^\varphi}{ 8\pi \varepsilon_p^\delta};\\
%
%
\!\!\!S^z_\Phi=\frac{e^4_z\Phi^2}{2\pi^2}\!\!\biggl(3F_1(\psi_z)-\frac{\psi^3_z}{\sqrt{1+\psi^2_z}}-\frac{\psi^2_z}{\sqrt{1+\psi^2_z}}\Delta_\Phi\biggr)
;\nonumber\\
S^z_\varphi=-\frac{e^4_z\Phi^2\psi^2_z}{2\pi^2\sqrt{1+\psi^2_z}}\Delta_\varphi; S^\zeta_\Phi=-\frac{e^4_\zeta\varphi^2\psi^2_\zeta}{2\pi^2\sqrt{1+\psi^2_\zeta}}\Delta_\Phi; \nonumber
\end{eqnarray}
\begin{eqnarray}
\label{s_zeta-delta}
\!\!\!\!\! S^\zeta_\varphi=\frac{e^4_\zeta\varphi^2}{2\pi^2}\!\!\biggl(3F_1(\psi_\zeta)-\frac{\psi^3_\zeta}{\sqrt{1+\psi^2_\zeta}}-
\frac{\psi^2_\zeta}{\sqrt{1+\psi^2_\zeta}}\Delta_\varphi\biggr);\nonumber\\
%
\!\!\!p_p^\delta=\frac{1}{\pi^2}\biggl(\frac{e_z^4\Phi^4\psi_z^4}{\sqrt{1+\psi_z^2}}
+\frac{e_\zeta^4\varphi^4\psi_\zeta^4}{\sqrt{1+\psi_\zeta^2}}\biggl)>0;\nonumber\\
%
\label{P^Phi}
\!\!\!P^\Phi=\frac{e^4_z\Phi^3}{2\pi^2}F_1(\psi_z)-p_p^\delta\Delta_\Phi;
P^\varphi=\frac{e^4_\zeta\varphi^3}{2\pi^2}F_1(\psi_\zeta)-p_p^\delta\Delta_\varphi.\nonumber
\end{eqnarray}
The coefficients \eqref{de_delta}, in turn, are used to determine macroscopic scalars $\delta\sigma^z$, $\delta\sigma^\zeta$, and $\delta p_p$, perturbations of scalar charge and pressure densities (see \cite{Ign_GC21_Un}).

Further, \eqref{gamma_ik} and \eqref{de_delta} yield the relations
\begin{equation}\label{gamma_33}
0<\frac{p^\delta_p}{\varepsilon^\delta_p}<1,\Rightarrow \frac{1}{3}<\gamma_{33}<\frac{4}{3}.
\end{equation}

Further, the coefficients $\gamma_{\alpha\beta}$ \eqref{gamma_ik}, on the one hand, are determined by the parameters of the model of scalar fields and charged fermions:
\begin{eqnarray}\label{parms}
\mathbf{p}:=[[\alpha,m,e_z,\pi_c],[\beta,\mathfrak{m},e_\zeta,\pi_f]];\\
\label{gamma_parms}
\gamma_{\alpha,\beta}=\gamma_{\alpha,\beta}(p,[\xi,\Phi,\varphi]).
\end{eqnarray}
On the other hand, the coefficients $\gamma_{\alpha\beta}$ depend on the base functions $\xi(t),\Phi(t),\varphi(t)$ of the unperturbed cosmological model, and therefore, in the end, are deter\-mined also the cosmological constant and the initial conditions:
\begin{eqnarray}\label{Params}
\mathbf{P}:=[\mathbf{p}\cup\Lambda]= [[\alpha,m,e_z,\pi_c],[\beta,\mathfrak{m},e_\zeta,\pi_f],\Lambda];\\
\label{inits}
\mathbf{I}:=[\Phi_0,Z_0,\varphi_0,z_0,\varrho],
\end{eqnarray}
where $[\xi(0)=0,\Phi(0)=\Phi_0,Z(0)=Z_0,\varphi(0)=\varphi_0,z(0)=z_0]$, $\varrho=\ pm 1$, so $H(0)=\varrho H_0$, $H_0$ is a non-negative solution of the Einstein equation \eqref{SurfEinst_M1} with respect to $H$.\footnote{See \cite{Ign_GC21_Un} for details.}

A necessary and sufficient condition for the non-trivial solvability of a system of linear homogeneous algebraic equations \eqref{AX=0} is that the determinant matrix $A$ vanishes, which leads to the \emph{dispersion equation},
which determines the dependence of the eikonal functions $u(t)$ on the wave number $n$: $u(n,t)$. Since $\mathrm{det}(\mathbf{A})$ contains only even powers of the eikonal function, the solutions of the dispersion equation
\begin{equation}\label{disp_eq0}
\mathrm{det}(\mathbb{A})\equiv -u^2\cdot \mathrm{det}(\mathbf{A}_3)=0
\end{equation}
are 4 pairs of symmetric solutions $\pm u(n,t)_k$, $k=\overline{1,4}$ corresponding to four pairs of \emph{perturbation modes}. The solutions of the dispersion equation\newline $\pm u_4=0$ correspond to the solution of the equation \eqref{AX=0} $\nu =0$ $\Rightarrow \lambda=-\mu$ \footnote{These modes are eliminated by admissible transformations of the metric (see Fig. \cite{Lifshitz}).}. The remaining three pairs of perturbation modes $\Phi, \varphi,\nu$, corresponding to the eikonal functions $\pm u_\alpha(n,t)$ $\alpha=\overline{1,3}$, are determined by a system of third-order linear equations
\begin{equation}\label{A3X=0}
 \mathbf{A}_3  \cdot
\left(\begin{array}{c}
\Phi\\ \varphi\\ \nu\\
\end{array}
\right)=0,
\end{equation}
where
\small{
\begin{eqnarray}\label{A_g}
\!\!\!\!\!\mathbf{A}_3=\left[\begin{array}{lll}
 1-x+\frac{\gamma_{11}}{n^2} & \frac{\gamma_{12}}{n^2} & \gamma_{13}\\[12pt]
\frac{\gamma_{21}}{n^2} & 1-x +\frac{\gamma_{22}}{n^2} & \gamma_{23}\\[12pt]
\frac{\gamma_{31}}{n^2} & \frac{\gamma_{32}}{n^2} & \gamma_{33}-x\\
\end{array}\right].
\end{eqnarray}}

In this case, in turn, the symmetric functions of the eikonal are determined by the third-order algebraic equation
\begin{equation}\label{det(A_3)=0}
\mathrm{det}(\mathbf{A}_3)=0
\end{equation}
with respect to the dimensionless function $x$
\begin{equation}\label{x}
x=\frac{u^2}{n^2}\sim 1; \;  u_\pm= \pm n \sqrt{x}.
\end{equation}
Since the equations \eqref{A3X=0} are sign-invariant $\pm u(n,t)$, the perturbations corresponding to the eikonal functions $\pm u(n,t)$ -- $\{\tilde{\ Phi}_\pm(n,t)$, $\tilde{\varphi}_\pm(n,t)$,
$\tilde{\nu}_\pm(n,t)\}$ must match:
\begin{eqnarray}\label{X_pm}
\tilde{\Phi}_-(n,t)=\tilde{\Phi}_+(n,t);\; \tilde{\varphi}_-(n,t)=\tilde{\varphi}_+(n,t);\nonumber\\ \tilde{\nu}_-(n,t)=\tilde{\nu}_+(n,t).
\end{eqnarray}

\subsection{Dispersion equation, modes and\newline perturbation masses\label{disp_eq}}
Solving the dispersion equation \eqref{det(A_3)=0} of the third order with respect to the dimensionless variable $x$ \eqref{x}, one can, of course, find its roots. At least one of the roots of this equation, for example
$x_3$ must be real, and the other two, for example $x_1, x_2$, must be real or complex conjugate. One can, of course, write out these solutions, but their form is extremely cumbersome, which hinders their analysis.
\subsubsection{Shortwave limit in the rigid WKB approximation}
Note, however, that in the rigid WKB approximation
\begin{equation}\label{n8}
n^2\to \infty ;\; \frac{n^2}{a^2}\gg \{m^2,\mathfrak{m}^2,\alpha\Phi^2,\beta\varphi^2\}
\end{equation}
solutions of the dispersion equation have the following asymptotics:
\begin{equation}\label{u_1_3(8)}
\left.x_{1,2}\right|_{n\to\infty}= 1; \; \left.x_{3}\right|_{n\to\infty}= \gamma_{33}.
\end{equation}
According to the equations \eqref{A3X=0} and the relation \eqref{A_g}, these asymptotics correspond to the following solutions:
\begin{eqnarray}
x_{1,2}:\ [\Phi,\varphi,\nu]_{1,2}=
[\mathrm{C}^\Phi_{1,2}(\eta)\mathrm{e}^{in(z\pm \eta)},\nonumber\\
\mathrm{C}^\varphi_{1,2}(\eta)\mathrm{e}^{in(z\pm \eta)},0], \nonumber\\
x_3:\ [\Phi,\varphi,\nu]_3=[-\gamma_{13},-\gamma_{23},1-\gamma_{33}]\times\nonumber\\
\mathrm{C}(\eta)\mathrm{e}^{in(z\pm \int \sqrt{\gamma_{33}(\eta)}d\eta)},\nonumber
\end{eqnarray}
where $\mathrm{C}^{\Phi,\varphi}_{1,2}(\eta),\mathrm{C}(\eta)$ are arbitrary slowly varying func\-tions\footnote{Their weak dependence on $ \eta$ determine the imaginary parts of the equations of the first WKB approximation \cite{STFI_20}.} Because due to \eqref{gamma_33} $\gamma_{33}>0$, all $X_\alpha$ oscillation modes at $n\to\infty$ are pairs of undamped retarded and advanced waves. Note that in the $X_{1,2}$ modes propagating at the speed of light, there are no pertur\-ba\-tions of the $\nu$ metric, while the phase velocity of propagation of pertur\-ba\-tions of the $X_3$ mode is equal to $v_f=\sqrt{\gamma_{33}}$.

\subsubsection{Solution of the dispersion equation and perturbation mode}
Let us turn to the solution of the cubic dispersion equation \eqref{det(A_3)=0} with respect to the dimensionless variable $x$ \eqref{x}, denoting the eikonal functions corresponding to the asymptotic solutions \eqref{u_1_3(8)} as
\begin{equation}\label{sol_u}
u^\pm_\alpha(n,\mathbf{p},[\xi,\Phi,\varphi])\equiv \pm u_\alpha(n,\eta)=\pm n\sqrt{x_\alpha},
\end{equation}
where $\mathbf{p}$ is an ordered set of \eqref{parms} parameters.
Passing to the cosmological time using the formula \eqref{ds0} in the \eqref{Eiconal} expressions and separating the real and imaginary parts in the eikonal functions, we obtain according to (see \cite{Ign22_I})
\begin{eqnarray}
i\int\limits_{\eta_0}^\eta u^\pm_\alpha d\eta= i\int\limits_0^t \omega^\pm_\alpha dt-\int\limits_0^t \gamma^\pm_\alpha dt,
\end{eqnarray}
where $\omega(t)$ and $\gamma(t)$ are local frequency and decrement\!\!/\!\! oscillation increment on the cosmological time scale $t$:
\begin{eqnarray}\label{g,o}
\omega^\pm_\alpha=\pm\mathrm{e}^{-\xi(t)}\mathrm{Re}\left(u^\pm_\alpha\right);\nonumber\\
\gamma^\pm_\alpha=\mp\mathrm{e}^{-\xi(t)}\mathrm{Im}\left(u^\pm_\alpha\right),
\end{eqnarray}
and $\tilde{\omega}^\pm_\alpha$ and $\tilde{\gamma}^\pm_\alpha$ are local frequency and decrement\!\! /\!\! oscillation increment on the scale of the time variable $\eta$.

In the general case, disturbances represent two groups of retarded and advanced waves propagating with a phase velocity
\begin{equation}\label{v_f}
v_f=\frac{\varpi}{n}\equiv a\frac{\omega_\pm}{n}
\end{equation}
with \emph{exponentially} decaying or growing amplitudes
\begin{equation}\label{instability}
\tilde{f}^-(\eta)\mathrm{e}^{ -\int\tilde{\gamma}(n,\eta)d\eta},\; \tilde{f}^+(\eta)\mathrm{e}^{+\int\tilde{\gamma}(n,\eta)d\eta}.
\end{equation}
The growing oscillation modes correspond to the instability of the homogeneous unperturbed state of the cosmological model. Further, according to \eqref{instability}, the amplitude of the growing disturbance mode at time $t$, \emph{growth factor of the disturbance amplitude}, is determined by the expression
\begin{equation}\label{chi}
\chi(t)=\int\limits_{t_1}^t \gamma(t)dt,
\end{equation}
where $t_1$ is the initial moment of instability occurrence. Let $t_2$ be the end time of the unstable phase, so that for $t>t_2$ $\gamma(t)=0$. Thus, during the development of instability on the interval $\Delta t=t_2-t_1$, the perturbation amplitude is fixed at $\tilde{f}^+(t)\exp(\chi_\infty)$, where
\begin{equation}\label{chi8}
\chi_\infty=\int\limits_{t_1}^{t_2} \gamma(t)dt.
\end{equation}
\subsection{BHS Formation Conditions and\newline Perturbation Mass Evolution}
For the formation of BHS -- (Black Hole Seeds) superheavy nuclei in scalar-gravitational perturbations in the early Universe, two conditions must be met simultaneously:
\begin{enumerate}\label{2Usl}
\item Exponential instability of some mode of perturbations with sufficiently large oscillation growth factor $\chi(t)$ \eqref{chi} to form a primordial black hole;
\item Possibility of reaching the mass of the primordial black hole on the order of $M_{bhs}$ \eqref{M_nc} at cosmological times of the order of $0.6\div1$ mlr. years.
\end{enumerate}

The evolution of the mass $M(t)$ of a black hole formed in a perturbation with a wave number $n$ is described by the differential equation \cite{Ign22_II}
\begin{equation}\label{dM/dt}
\!\!\frac{dM}{dt}+\frac{1}{\kappa M^2}=\frac{4\pi}{n^3}H^3(t)\mathrm{e}^{3\xi(t)}[3+2(\Omega(t)-1)].
\end{equation}
with the initial condition:
\begin{equation}\label{IC}
M(t_g)=M(t_g,n).
\end{equation}
The equations \eqref{dM/dt} -- \eqref{IC} have the following notation: $\kappa=15\cdot2^{10}\pi$, $M(n,t)$ is the reduced effective mass of the perturbation region enclosed in a sphere with a radius equal to the length of the perturbation $\lambda(t)=a(t)/n\equiv \mathrm{e}^{\xi(t)} /n$:
\begin{equation}\label{M_n}
M(n,t)=\frac{4\pi}{3}\lambda^3(t)\mathcal{E}_{eff}(t)=\frac{4\pi}{n^3}H^2(t)\mathrm{e}^{3\xi(t)},
\end{equation}
and $t_g$ is the black hole formation time in the $n$ perturbation mode, which is determined from the equation
\begin{equation}\label{t_n}
\lambda(t)=2M(n,t)\Rightarrow\displaystyle 8\pi\frac{H^2(t_g)}{n^2}\mathrm{e}^{2\xi(t_g)}=1.
\end{equation}
The \eqref{dM/dt} equation takes into account two processes that change the mass of a black hole: the cosmological expansion factor and the mass loss due to black hole evaporation. Note that the coefficients of the nonlinear inhomogeneous equation
\eqref{dM/dt} $\xi(t)$, $H(t)$ and $\Omega(t)$, in turn, are determined by numerical solutions of the nonlinear system of background equations \eqref{dxi/dt-dPhi_Phi} -- \eqref{SurfEinst_M1}.

\section{Numerical modeling}
\subsection{Parameters and initial conditions}
Turning to numerical simulation, further, to reduce the letter, we will specify a set of fundamental parameters of the \MI model using an ordered list (see \cite{GC_21_3})
\[\mathbf{P}=[[\alpha,m,e,m_c,\pi_c],[\beta,\mu,\epsilon,m_f,\pi_f],\Lambda]\]
and the initial conditions by an ordered list
\[\mathbf{I}=[\Phi_0,Z_0,\varphi_0,z_0,\kappa],\]
where $\kappa=\pm1$, and the value $\kappa=+1$ corresponds to the non-negative initial value of the Hubble parameter $H_0=H_+\geqslant0$, and the value $\kappa=-1$ corresponds to the negative initial value of the Hubble parameter $H_0 =H_-<0$. Here, using the autonomy of the dynamical system, we set $\xi(0)=0$ everywhere. Thus, the \MI model is determined by 11 fundamental para\-me\-ters and 5 initial conditions.
\subsection{Singular points of the background vacuum scalar doublet}
In what follows, we will need the coordinates of \emph{singular points of the vacuum scalar Higgs doublet of the back\-ground metric} in the 3-dimensional subspace $\mathbb{R}_3\{H,\Phi,\varphi\}$ of the 5-dimensional phase space of the corresponding dynamical system $\mathbb{R}_5=\{H,\Phi,Z,\varphi,z\}$ \cite{Ignat21_TMP}, since these points largely determine the dynamics of the \MI back\-ground model.\footnote{In In what follows, a vacuum scalar doublet will be referred to as a cosmological model in which there is no matter other than scalar fields. This model is obtained by passing to the limit from the \MI\ model at $\pi_c=\pi_f=0$.}
For positive fundamental parameters $\alpha,\beta,\Lambda$, there are 18 such singular points. These are, firstly, two symmetrical points located on the $OH$ axis --\footnote{The first derivatives of the scalar potentials $Z=z=0$ vanish at all singular points.}
\begin{equation}\label{M_pm}
M_{0,0}^\pm\biggl(0,0,\pm\sqrt{\frac{\Lambda_0}{3}}\biggr),
\end{equation}
secondly, 8 $M^\pm_{\pm1,\pm1}$ points symmetric with respect to the plane $\{\Phi,\varphi\}$ --
\begin{eqnarray}\label{M_1,1}
M^\pm_{\pm1,\pm1}\biggl(\pm\frac{m}{\sqrt{\alpha}},\pm\frac{\mathfrak{m}}{\sqrt{\beta}}, \pm \sqrt{\frac{\Lambda}{3}}\biggr),
\end{eqnarray}
and, thirdly, 8 points $M^\pm_{0,\pm1}$ and $M^\pm_{\pm1,0}$ symmetric with respect to the plane $\{\Phi,\varphi\}$:
\begin{eqnarray}\label{M_0,1}
M^\pm_{0,\pm1}\biggl(0,\pm\frac{\mathfrak{m}}{\sqrt{\beta}}, \pm \frac{1}{\sqrt{3}}\sqrt{\Lambda_0+\frac{m^4}{\alpha}} \biggr),\\
\label{M_1,0}
M^\pm_{\pm1, 0}\biggl(\pm\frac{m}{\sqrt{\alpha}},0, \pm \frac{1}{\sqrt{3}}\sqrt{\Lambda_0+\frac{\mathfrak{m}^4}{\beta}} \biggr).
\end{eqnarray}

However, of the above singular points of the vacuum scalar doublet of the background metric, only two points $M^+_{0,\pm1}$ are attracting in the full phase space $\mathbb{R}_5$. The two-dimensional subspaces $\Sigma_\Phi=\{\Phi,H\}$ and $\Sigma_\varphi=\{\varphi,H\}$ have attraction points $M^+_{0,0}$ and $ M^+_{\pm1,\pm1}$, res\-pec\-ti\-vely, but these points turn out to be unstable in the complete phase space $\mathbb{R}_5$.

\subsection{Numerical Simulation Example}
Let us consider the behavior of the perturbation evolution model in the case of the following set of fundamental parameters corresponding to small values of the scalar charges $e_z$, $E_\zeta$ and small values of the cosmological constant:
\begin{equation}\label{params1}
\mathbf{P}=[[1,1,10^{-5},0.1],[1,1,10^{-5},0.1],10^{-5}]
\end{equation}
In this case, the singular points of the vacuum doublet \eqref{M_pm} -- \eqref{M_1,0} correspond to the following values of $[H,\Phi,\varphi]$:
\begin{equation} \label{sing_ik}
\begin{array}{ccccc}
M   & M^\pm_{0,0} & M^\pm_{\pm1,\pm1} & M^\pm_{0,\pm1} & M^\pm_{\pm1,0}\\[2pt]
\hline\\[-8pt]
H_0 & \pm0.1826 & \pm0.4082 & \pm0.2887 & \pm0.2887;\\
\Phi_0 & 0 & \pm1 & 0 &  \pm1;\\
\varphi_0 & 0 & \pm1 & \pm1 & 0.\\
\end{array}
\end{equation}
In this case, the stable singular points \eqref{M_0,1} have coordinates:
\begin{equation}\label{sing_1}
M^+_{0,\pm1}=[0,\pm1,\pm0.2887],
\end{equation}
the other singular points are unstable.
\Fig{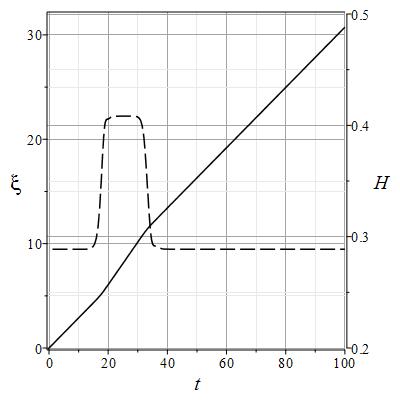}{6}{\label{ignatev1}Evolution of the scale function $\xi(t)$ (solid lines) and the Hubble parameter $H(t)$ (dashed lines) in the case of parameters \eqref{params1}.}
\Fig{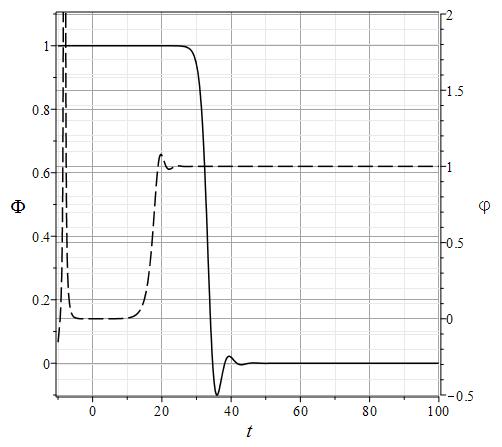}{6}{\label{ignatev2}Evolution of scalar potentials: solid lines - $\Phi(t)$, dashed lines - $\varphi(t)$ in case of \eqref{params1} parameters.}

Figure \ref{ignatev1} shows the evolution of the scale function and the Hubble parameter for these fundamental pa\-ra\-meter values, and Figure \ref{ignatev2} shows the evolution of scalar potentials. As can be seen from these figures, the unper\-turbed dynamical system evolves to a stable vacuum singular point $M^+_{0,1}$, and in the process of evolution the system lingers for some time in the neighborhood of unstable singular points $M^+_{1,0} $ and $M^+_{1,1}$.

Next, Figure \ref{ignatev3} -- \ref{ignatev5} shows the graphs of the evolution of frequencies $\omega(t)=\Re(u(t))$ and increments $\gamma(t)=-\Im(u (t)$ oscillations of the $2,3,5$ modes (the graphs of the $1,4,6$ modes are obtained by reflecting these graphs from the x-axis). second, the $u_{1,2}$ modes are stable, as are the $u_{4,6}$ modes, while the $u_{3,5}$ modes are unstable. graphs with graphs on Figure \ref{ignatev1} -- \ref{ignatev2}, one can notice that the indicated instability occurs near the point $M^+_{1,0}$ of unstable equilibrium of an unperturbed dynamical system with vacuum Higgs scalar fields. , it can be seen that the amplitude growth factor \eqref{chi8} for the $u_3$ mode is about $\chi_\infty\approx 20$, and for the $u_5$ mode it is about $\chi_\infty\approx35$.

\Fig{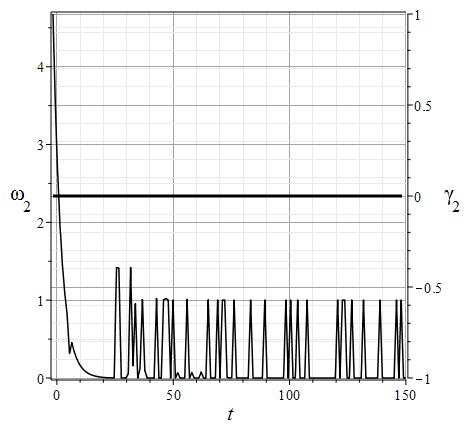}{6}{\label{ignatev3}Evolution of the oscillation frequency $\omega(t)$ and the oscillation increment $\gamma(t)=-\Im(u(t)$ (thick line) for the $u_2$ mode in the case of the parameters \eqref{params1} and $n= 3$.}
\Fig{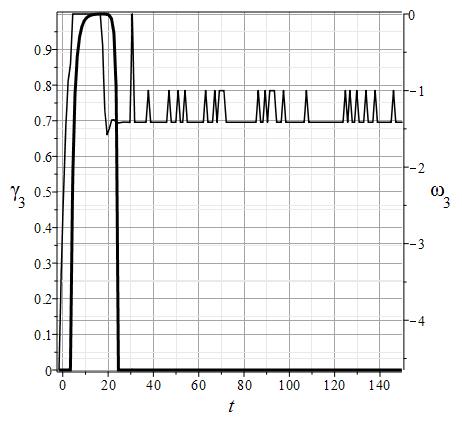}{6}{\label{ignatev4}The evolution of the oscillation frequency $\omega(t)$ and the oscillation increment $\gamma(t)=-\Im(u(t)$ (thick line) for the $u_3$ mode in the case of the parameters \eqref{params1} and $n= 3$.}
\Fig{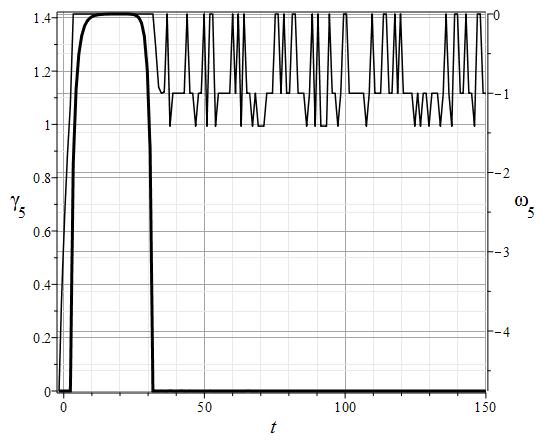}{6}{\label{ignatev5}Evolution of the oscillation frequency $\omega(t)$ and the oscillation increment $\gamma(t)=-\Im(u(t)$ (bold line) for the $u_5$ mode in the case of the parameters \eqref{params1} and $n= 3$.}

Thus, the first condition of Section \ref{2Usl} is fulfilled. Figure \ref{ignatev6} shows the results of numerical integration of the black hole mass evolution equation \eqref{dM/dt}.
\Fig{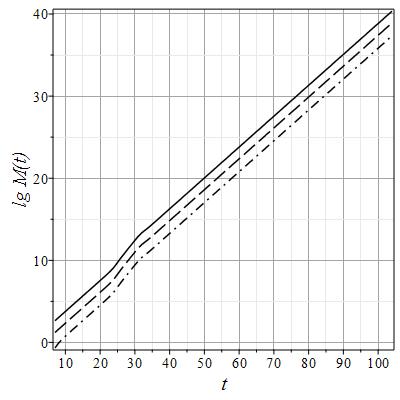}{6}{\label{ignatev6}Black hole mass evolution in the case of \eqref{params1} parameters: solid line -- $n=1$, dashed line -- $n=3$, dash-dotted line -- $n=10$.}

Thus, the second condition of Section \ref{2Usl} is not satisfied - the black holes formed at the instability stage $t\in[5,30]$ reach maximum values of the order of $10^{13}m_{\mathrm{pl) at this stage }}$, which is incomparably small compared to the required masses $M_{bhs}\sim10^{42}\div10^{44}m_{\mathrm{pl}}$.
\subsection{Another example of numerical\newline modeling}
Consider the behavior of the perturbation evolution model in another case of a set of fundamental parameters, in which we increase the scalar charges while decreasing the initial Fermi momenta:
\begin{equation}\label{params2}
\mathbf{P}=[[1,1,10^{-4},10^{-2}],[1,1,10^{-3},10^{-3}],10^{-5}]
\end{equation}
In this case, the singular points of the vacuum doublet \eqref{M_pm} -- \eqref{M_1,0} have the same coordinates as in the previous case -- \eqref{sing_ik}, however, the behavior of the \MI\ model in this case is significantly more difficult. Figure \ref{ignatev7} shows the evolution of the scale function and the Hubble parameter for these fundamental parameter values, and Figure \ref{ignatev8} shows the evolution of scalar potentials. As can be seen from these figures, the unperturbed dynamical system evolves to a stable vacuum singular point $M^+_{0,1}$, and in the process of evolution the system lingers for some time in the neighborhood of unstable singular points $M^+_{1,0} $ and $M^+_{1,1}$.
\Fig{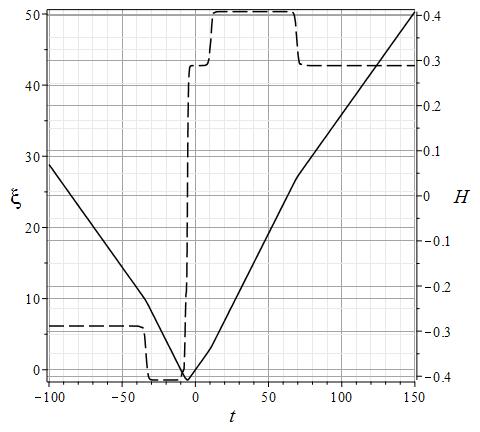}{6}{\label{ignatev7}Evolution of the scale function $\xi(t)$ (solid lines) and the Hubble parameter $H(t)$ (dashed lines) in the case of \eqref{params2} parameters.}
\Fig{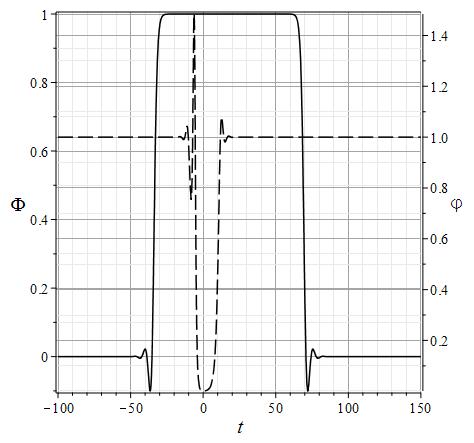}{6}{\label{ignatev8}Evolution of scalar potentials: solid lines - $\Phi(t)$, dashed lines - $\varphi(t)$ in case of \eqref{params2} parameters.}

Figures \ref{ignatev9} and \ref{ignatev10} show the evolution of frequencies and increments of oscillations for the $u_3,u_5$ modes, the oscillation increment for the $u_1$ mode is zero. On these plots, we observe a reduction in the $u_3$ mode instability interval by about a factor of two while simultaneously increasing the $u_5$ mode instability interval by a factor of two compared to the previous example, which gives $\chi_\infty\approx 70$ for the perturbation growth factor. Note that in this case the oscillation amplitude grows by a factor of $\exp(\chi_\infty)\approx 2.5\cdot10^{30}$ times!

\Fig{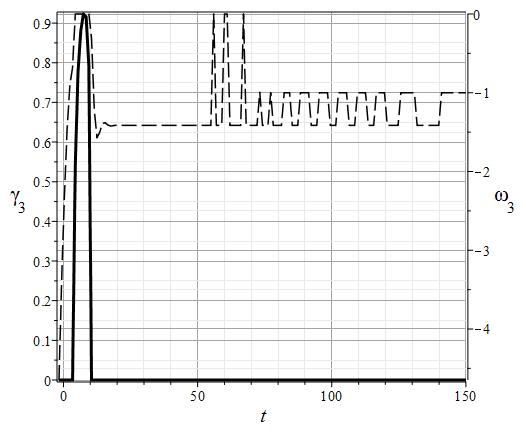}{6}{\label{ignatev9}Evolution of the oscillation frequency $\omega(t)$ and the oscillation increment $\gamma(t)=-\Im(u(t)$ (bold line) for the $u_3$ mode in the case of the parameters \eqref{params2} and $n= 3$.}
\Fig{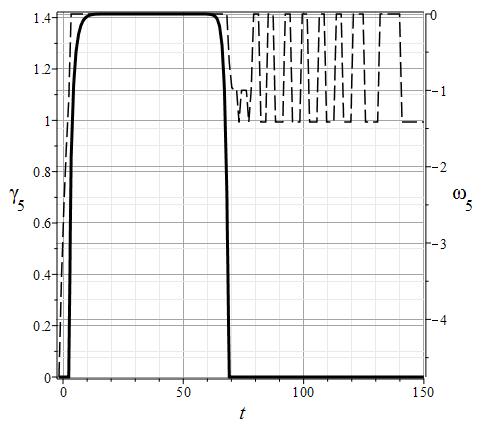}{6}{\label{ignatev10}Evolution of the oscillation frequency $\omega(t)$ and the oscillation increment $\gamma(t)=-\Im(u(t)$ (bold line) for the $u_5$ mode in the case of the parameters \eqref{params2} and $n= 3$.}

Thus, condition 1 of Section \ref{2Usl} in the case under consideration is satisfied with a huge margin, which allows us to quickly bring even very small perturbations to the nonlinear stage.

Let us now turn to the analysis of the fulfillment of condition 2 of Section \ref{2Usl}. Figure \ref{ignatev11} shows the evolution of the reduced mass of a spherical perturbation corresponding to wavenumber $n$. The gray stripe in this picture
the mass region HN \eqref{M_nc} is shown. Thus, in this case, all $M(t)$ curves ``reach out'' to the band \eqref{M_nc}, i.e., both conditions of the \ref{2Usl} section are satisfied, which ensures the BHS formation process in a fairly wide range of wave numbers $n=1\div10$.
\Fig{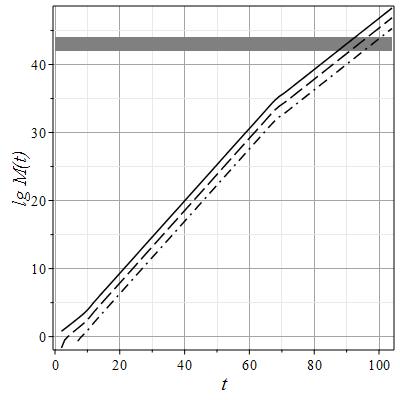}{6}{\label{ignatev11}Black hole mass evolution in the case of \eqref{params2} parameters: solid line -- $n=1$, dashed line -- $n=3$, line length -- $n=10$.}
\section{WKB - approximation for field equations without sources: stabi\-lity of perturbations}
For a better understanding of the influence of a phantom scalar field on the process of development of gravitational-scalar instability, we study the stability of short-wave perturbations of free scalar fields.
\subsection{WKB solution for eikonal functions}
So, consider the perturbed field equations without sources, representing perturbations in the form \eqref{dF-drho-du} and putting in \eqref{Box(Phi)=sigma_z} -- \eqref{Box(varphi)=sigma_zeta} $\sigma ^z=0$ and $\sigma^\zeta=0$
\begin{eqnarray}\label{Eq_dPhi}
\delta\Phi''+2\frac{a'}{a}\delta\Phi'+\bigl[n^2+a^2(m^2-3\alpha\Phi^2)\bigr]\delta\Phi \nonumber\\
+\frac{1}{2}\Phi'\mu'=0;\\
\label{Eq_dvarphi}
\delta\varphi''+2\frac{a'}{a}\delta\varphi'+\bigl[n^2-a^2(\mathfrak{m}^2-3\beta\varphi^2)\bigr]\delta\varphi \nonumber\\
+\frac{1}{2}\varphi'\tilde{\mu}'=0.
\end{eqnarray}
Representing the perturbations in the eikonal form \eqref{Eiconal} into the equations \eqref{Eq_dPhi} -- \eqref{Eq_dvarphi}, we obtain equations for the perturbations $\delta\tilde{\Phi}$, $\delta\tilde{\varphi} $, $\tilde{\mu}$ and eikonal $u$:
\begin{eqnarray}\label{Eq_dPhi_WKB}
[n^2-u^2+a^2(m^2-3\alpha\Phi^2)]\delta\tilde{\Phi}\\
+i\bigl[2u\delta\tilde{\Phi}'+2\frac{a'}{a}u\delta\tilde{\Phi}+u'\delta\tilde{\Phi}+\frac{1}{2}\Phi'u\tilde{\mu}\bigr]\nonumber\\
+\biggl[\delta\tilde{\Phi}''+2\frac{a'}{a}\delta\tilde{\Phi}'+\frac{1}{2}\Phi'\tilde{\mu}'\biggr] =0;\nonumber
\end{eqnarray}
\begin{eqnarray}\label{Eq_dvarphi_WKB}
[n^2-u^2-a^2(\mathfrak{m}^2-3\beta\varphi^2)]\delta\tilde{\varphi}\\
+i\bigl[2u\delta\tilde{\varphi}'+2\frac{a'}{a}u\delta\tilde{\varphi}+u'\delta\tilde{\varphi}+\frac{1}{2}\varphi'u \tilde{\mu}\bigr]\nonumber\\
+\biggl[\delta\tilde{\varphi}''+2\frac{a'}{a}\delta\tilde{\varphi}'+\frac{1}{2}\varphi'\tilde{\mu}'\biggr]=0.\nonumber
\end{eqnarray}
The square brackets of the left parts of the equations \eqref{Eq_dPhi_WKB} -- \eqref{Eq_dvarphi_WKB} contain terms of the zero, first and second orders of the WKB approximation, in order.

Thus, in the zero WKB approximation for the eikonal functions, we obtain from \eqref{Eq_dPhi_WKB} -- \eqref{Eq_dvarphi_WKB}
\begin{eqnarray}\label{u_c}
u^\pm_c=\pm u_c\equiv\pm\sqrt{n^2+a^2(m^2-3\alpha\Phi^2)};\\
\label{u_f}
u^\pm_f=\pm u_f\equiv\pm\sqrt{n^2-a^2(\mathfrak{m}^2-3\beta\varphi^2)},
\end{eqnarray}
where $u_c$ and $u_f$ are the eikonal functions for the classical and phantom fields, respectively. Note that in the shortwave limit \eqref{n8} the solutions \eqref{u_c} -- \eqref{u_f} coincide with the asymptotic solutions \eqref{u_1_3(8)} corresponding to the values of $x_{1,2}$.

\subsection{Areas of sustainability}
Thus, at low absolute value potentials of the background scalar fields
\begin{eqnarray}\label{F<}
\Phi^2<\frac{m^2}{3\alpha};\quad \varphi^2<\frac{\mathfrak{m}^2}{3\beta}
\end{eqnarray}
the expression under the right-hand side radical \eqref{u_c} increases with increasing scale factor, while the expression under the right-hand side radical \eqref{u_f}, on the contrary, decreases. In this case, provided
\begin{equation}\label{a^2>=f}
n^2<a^2(t)(\mathfrak{m}^2-3\beta\varphi^2(t))
\end{equation}
the phantom field eikonal functions \eqref{u_f} become purely imaginary, which corresponds to standing waves, i.e., instability of phantom field perturbations. In this case, the classical field eikonal functions \eqref{u_c} remain real, which corresponds to retarded and advanced waves, i.e., stable perturbations of the classical field.

For sufficiently large absolute values of the potentials of the background scalar fields
\begin{eqnarray}\label{F>}
\Phi^2<\frac{m^2}{3\alpha};\quad \varphi^2<\frac{\mathfrak{m}^2}{3\beta}
\end{eqnarray}
the expression under the right side radical \eqref{u_c} decreases with increasing scale factor, while the expression under the right side radical \eqref{u_f}, on the contrary, increases. In this case, provided
\begin{equation}\label{a^2>=c}
n^2<a^2(t)(m^2-3\alpha\Phi^2(t))
\end{equation}
the classical field eikonal functions \eqref{u_c} become purely imaginary, which corresponds to standing waves, i.e., the instability of classical field perturbations. In this case, the phantom field eikonal functions \eqref{u_f} remain valid, which corresponds to retarded and advanced waves, i.e., stable perturbations of the phantom field.

\subsection{Shortwave Stability at Vacuum\newline Equilibrium Points}
In what follows, we will need the coordinates of stable (attractive) singular points of the dynamical system of the unperturbed cosmological model with \emph{vacuum} asymmetric scalar Higgs doublet in the three-dimensional phase subspace of the phase space of the corresponding 5-dimensional dynamical system$\mathbb{R}_3:$ $ [\Phi,\varphi,H,Z=0,z=0]$\footnote{This system is obtained from the dynamic system \eqref{dxi/dt-dPhi_Phi} -- \eqref{dzZ/dt_M1} by removing the variable $\xi $ with parameters $\pi_c=\pi_f=0$.}. According to \cite{Ignat21_TMP}, there are only 2 such symmetrical stable equilibrium points:
\begin{equation}\label{sing_point}
\!\!\!M^+_{0,\pm1}: \Phi_0=0;\; \varphi_0=\pm \frac{\mathfrak{m}}{\sqrt{\beta}};\; H_0=\frac{1}{\sqrt{3}}\sqrt{\Lambda+\frac{\mathfrak{m}^4}{4\beta}}.
\end{equation}
At these points, the eikonal functions are equal:
\begin{eqnarray}\label{u^0_c}
u^\pm_c=\pm\sqrt{n^2+a^2m^2},\quad (M^+_{0,\pm1});\\
\label{u^0_f}
u^\pm_f=\pm\sqrt{n^2+2a^2\mathfrak{m}^2},\quad (M^+_{0,\pm1}).
\end{eqnarray}
Thus, at the points of stable equilibrium of the dynamical system of the unperturbed cosmological model with \emph{vacuum} asymmetric scalar Higgs doublet, the eikonal functions for perturbations of both classical and phantom fields are real and represent pairs of retarded and advanced waves, i.e., all short-wavelength perturbations are stable. Note that the stable equilibrium point \eqref{sing_point} corresponds to the inflationary expansion mode $\Omega_0=1$.

In the case of a vacuum classical scalar singlet, there is only one stable equilibrium point $M^+_0(0,0,\sqrt{\Lambda/3})$. The eikonal functions for perturbations in the vicinity of this point \eqref{u_c} are also real, i.e., the perturbations are stable.

To prevent possible misunderstandings, let us\newline comment on the result obtained in more detail. Indeed, according to the qualitative theory of dynamical systems, the points $M^+_{0,\pm1}$ and $M^+_0$ are points of stable equi\-lib\-rium, but -- with respect to changing the initial conditions \emph{of a homogeneous and isotropic cosmological model} , i.e., with respect to perturbations of the metric and scalar fields independent of the coordinates $\delta g^i_k(t)$, $\delta\Phi(t)$, $\delta\varphi(t)$. Such perturbations correspond to the limit $n\to0$. As shown above, and in the shortwave limit \eqref{Eiconal} $n\eta\gg1$, perturbations of free scalar fields are stable near stable equilibrium points. As shown above, and in the shortwave limit \eqref{Eiconal} $n\eta\gg1$, perturbations free scalar fields are stable near stable equilibrium points of the homogeneous cosmological model.
\subsection{Solutions of First Order Equations}
Note that the eikonal functions $u^\pm_c$ \eqref{u_c} and $u^\pm_f$ \eqref{u_f} thus obtained must be further substituted into the equations of the first-order WKB approximation, whence we obtain
\begin{eqnarray}\label{Eq_dPhi_WKB1}
2u^\pm_c\delta\tilde{\Phi}'+2\frac{a'}{a}u^\pm_c\delta\tilde{\Phi}+(u^\pm_c)'\delta\tilde{\Phi}+\frac{1}{2}\Phi'u^\pm_c\tilde{\mu}=0;\\
\label{Eq_dvarphi_WKB1}
2u^\pm_f\delta\tilde{\varphi}'+2\frac{a'}{a}u^\pm_f\delta\tilde{\varphi}+(u^\pm_f)'\delta\tilde{\varphi}+\frac{1}{2}\varphi'u^\pm_f \tilde{\mu}=0.
\end{eqnarray}
For a given amplitude of gravitational perturbations $\tilde{\mu}(\eta)$, the equations \eqref{Eq_dPhi_WKB1} -- \eqref{Eq_dvarphi_WKB1} represent a system of independent inhomogeneous linear differential equations with respect to the amplitudes $\delta\tilde{\Phi}(\ eta)$ and $\delta\tilde{\varphi}(\eta)$. It is not difficult to find general solutions in quadratures of these equations
\begin{eqnarray}\label{Sol_dPhi_WKB1}
\delta\tilde{\Phi}=\frac{1}{a\sqrt{u_c}}\biggl[C_c-\frac{1}{2}\int\Phi'\tilde{\mu}a\sqrt{u_c}d\eta \biggr];\\
\label{Sol_dvarphi_WKB1}
\delta\tilde{\varphi}=\frac{1}{a\sqrt{u_f}}\biggl[C_f-\frac{1}{2}\int\varphi'\tilde{\mu}a\sqrt{u_f}d\eta \biggr],
\end{eqnarray}
where $C_c,\ C_f$ are arbitrary constants. In particular, on stable background solutions \eqref{sing_point} $\Phi'=\varphi'=0$ the integrals in the solutions \eqref{Sol_dPhi_WKB1} -- \eqref{Sol_dvarphi_WKB1} vanish,
and we'll get:
\begin{eqnarray}\label{Sol_dPhi_WKB1_sing}
\delta\tilde{\Phi}=\frac{C_c}{a\sqrt{u_c}},\; (\Phi=\mathrm{Const});\\
\label{Sol_dvarphi_WKB1-sing}
\delta\tilde{\varphi}=\frac{C_f}{a\sqrt{u_f}},\; (\varphi=\mathrm{Const}).
\end{eqnarray}
\subsection{Numerical modeling}
Further, the inequalities \eqref{a^2>=f} and \eqref{F>} are determined by the potentials of the scalar fields, which, in turn, depend on time, $\Phi(t)$ and $\varphi(t)$ , unless the dynamical system is at points of stable equilibrium. Therefore, the stability of short-wavelength perturbations of these fields also depends on time and a specific cosmological scenario that determines the time dependence of scalar potentials. Consider a specific numerical example of the evolution of free oscillations of scalar fields in comparison with the evolution of oscillations in a system of charged fermions:

\begin{equation}\label{Par_free}
\mathbf{P}=[[1,1,1,0.1],[1,1,1,0.1],0.1]
\end{equation}
In this case, the singular points of the vacuum doublet correspond to the following values of $H,\Phi,\varphi$:
\begin{equation}\label{SINGS}
\hspace{-12pt}\begin{array}{ccccc}
M   & M^\pm_{0,0} & M^\pm_{\pm1,\pm1} & M^\pm_{0,\pm1} & M^\pm_{\pm1,0}\\[2pt]
\hline\\[-8pt]
H_0 & \pm0.1826 & \pm0.4472 & \pm0.3416 & \pm0.34156;\\
\Phi_0 & 0 & \pm1 & 0 &  \pm1;\\
\varphi_0 & 0 & \pm1 & \pm1 & 0.\\
\end{array}
\end{equation}

The Figures \ref{ignatev12} and \ref{ignatev13} show the evolution of the basic functions of the \MI\ model with the \eqref{Par_free} parameters.
\Fig{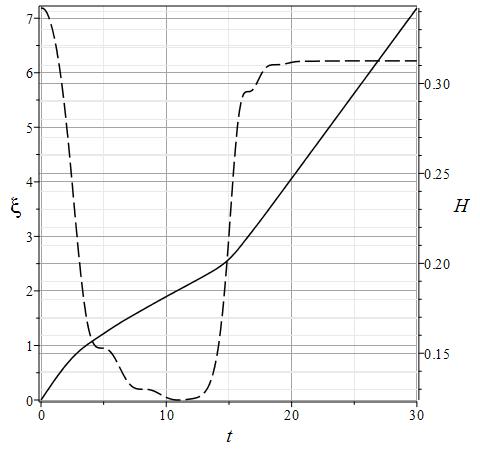}{6}{\label{ignatev12}Evolution of the scale function $\xi(t)$ (solid lines) and the Hubble parameter $H(t)$ (dashed line) in the \MI\ model in the case of parameters \eqref{Par_free}.}
\Fig{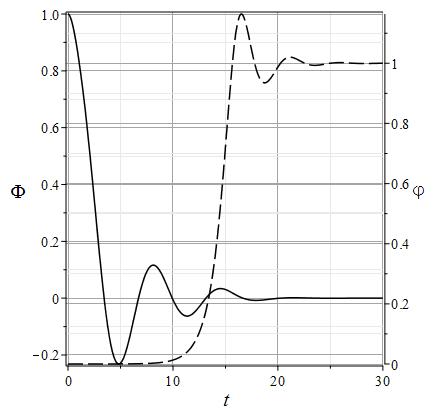}{6}{\label{ignatev13}Evolution of scalar potentials $\Phi(t)$ (solid lines) and $\varphi(t)$ (dashed line) in the \MI\ model in the case of parameters \eqref{Par_free}.}
Comparison of the graphs in these figures with the coordinates of the singular points \eqref{SINGS} leads to the conclusion that the cosmological model starts from a conditionally stable (in the two-dimensional direction) singular point $ M^+_{+1,+1}$, then passes through the unstable singular point $M^+_{0,0}$ and finally hits the stable singular point $M^+_{0,+1}$.

Further, Figure \ref{ignatev14} -- \ref{ignatev15} shows the results of numerical simulation of the oscillation increment $\gamma(t)$ for free classical and phantom fields (dashed lines) in comparison with similar results for the \MI\ ( solid lines) for the background model parameters \eqref{Par_free}, and Figure \ref{ignatev16} shows the results for the $u_{5,6}$ modes, which are absent in the free vibration model.
\Fig{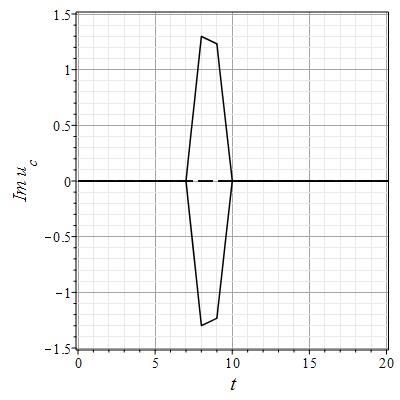}{6}{\label{ignatev14}Evolution of the oscillation increment $\gamma_{1,2}(t)=-\Im(u_{1,2}(t)$ in the \MI\ model (solid lines) and the increment of free oscillations of the phantom field $\gamma_c(t) $ (dashed line) in case of \eqref{Par_free} parameters.}
\Fig{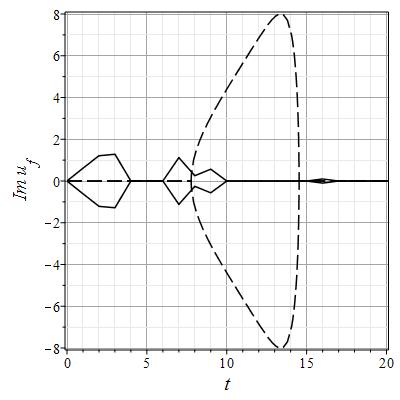}{6}{\label{ignatev15}Evolution of the oscillation increment $\gamma_{3,4}(t)=-\Im(u_{3,4}(t)$ in the \MI\ model (solid lines) and the increment of free oscillations of the phantom field $\gamma_c(t) $ (dashed line) in case of \eqref{Par_free} parameters.}
\Fig{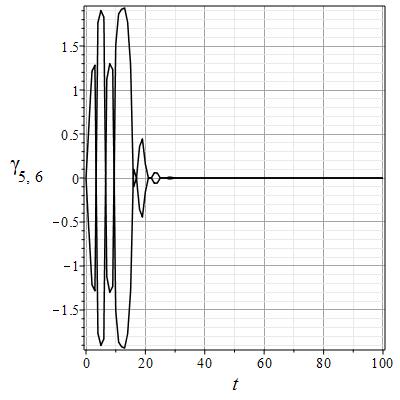}{6}{\label{ignatev16}The evolution of the oscillation increment $\gamma_{5,6}(t)=-\Im(u_{5,6}(t)$ in the \MI\ model in the case of parameters \eqref{Par_free}.}

Analyzing the graphs on Figure \ref{ignatev14} -- \ref{ignatev15}, one can notice, firstly, that for free fields the instabilities of perturbations also arise near the unstable points of the unperturbed vacuum doublet. Secondly, classical field perturbations turn out to be stable ($\gamma_c=0$), and the strong instability of phantom field perturbations is suppressed by charged fermions.
\section{Conclusion}
Let us briefly list the main results of the article.\\
\stroka{A mathematical model is constructed for the evolu\-tion of short-wavelength perturbations of a cosmological model based on a two-component system of degenerate scalarly charged fermions interacting through scalar classical and phantom Higgs fields. At the same time, in contrast to previous articles, the condition of smallness of scalar charges is removed.
}
\stroka{A numerical model of the evolution of short-wave disturbances is constructed, and the dispersion equation for the disturbance modes is solved by numerical methods.
}
\stroka{Apparently, in the presence of a phantom component, three of the four types of instabilities found in the one-field model do not arise: broadband, quasi-periodic, and aperiodic \cite{Ign22_I}.
}
\stroka{The instabilities have a narrow-band character and arise in the region of unstable points of an asymmetric vacuum doublet.
}
\stroka{In this case, the $\{1,2\}$ perturbation modes turn out to be stable and correspond to undamped waves, the $\{3,4\}$ modes have a narrow burst of instability at the earliest stages of evolution, the $\{5, 6\}$ are unstable on a sufficiently long time interval. It is these modes that correspond to collective oscillations of scalar and gravitational fields.
}
\stroka{Numerical modeling of the evolution of the pertur\-ba\-tion mass based on the previously formulated balance equation \cite{Ign22_II} showed the possibility of achieving the required BHS mass \eqref{M_nc} in this model for sufficiently large values of scalar charges $10^{-4}\div10{-3 }$.
}
\stroka{The evolution of short-wave perturbations of free scalar fields is studied. It is shown that at the points of stable equilibrium of the asymmetric vacuum doublet, the pertur\-ba\-tions of both the classical scalar and phantom fields are stable.
}
\stroka{Numerical simulation of perturbations of free fields in the \MI\ model in comparison with perturbations in a medium of charged fermions revealed the effect of suppressing the strong instability of perturbations of a free phantom field.
}

Thus, the results of the article once again showed the promise of the gravitational-scalar instability mechanism for building BHS models. Note that in order to build a sufficiently complete model of BHS formation, it is necessary to solve two fundamentally important tasks: \\
\noindent$\bullet$ Investigate the evolution of individual physical compo\-nents in unstable perturbations: fermion density and velocity perturbations, scalar field perturbations, and metrics. In particular, to clarify the question of the possible separation of these perturbations.\\
\noindent$\bullet$ Investigate the evolution of spherical perturbations in the cosmological model \MI.\\
We intend to study these problems in subsequent publi\-ca\-tions.

\subsection*{Funding}

  This article was supported by the Academic Strategic Leadership Program of Kazan Federal University.


%
%
\end{document}